\documentclass[aps,prb,reprint,groupedaddress]{revtex4-1}

\usepackage{amsmath}
\usepackage{graphicx}
\usepackage[caption=false,font=normalsize,labelfont=sf,textfont=sf]{subfig}

\begin{document}

\title{Effect of dipolar interactions on cavity magnon-polaritons}

\author{Antoine Morin}
\author{Christian Lacroix}
\author{David M\'enard}
\affiliation{Department of Engineering Physics, Polytechnique Montr\'eal, Montr\'eal, Qc}

\date{\today}

\begin{abstract}
The strong photon-magnon coupling between an electromagnetic cavity and two yttrium iron garnet (YIG) spheres has been investigated in the context of a strong mutual dipolar interaction between the spheres. A decrease in the coupling strength between the YIG spheres and the electromagnetic cavity is observed, along with an increase of the total magnetic losses, as the distance between the spheres is decreased. A model of inhomogeneous broadening of the ferromagnetic resonance linewidth, partly mitigated by the dipolar narrowing effect, reproduces the reduction in the coupling strength observed experimentally. These findings have important implications for the understanding of strongly coupled photon-magnon system involving densely packed magnetic objects, such as ferromagnetic nanowires arrays, in which the total coupling strength with an electromagnetic cavity might become limited due to mutual dipolar interactions.
\end{abstract}

\pacs{}

\maketitle

\section{\label{sec:int}INTRODUCTION}
Following recent works on the strong coupling between a magnonic mode of a ferromagnetic sample and a photonic mode of a microwave cavity\cite{soykal2010strong,huebl2013high}, also called cavity magnon-polaritons\cite{cao2015exchange,yao2015theory}, much interest emerged in the scientific community to exploit the phenomenon as a mean to develop novel information transfer technologies\cite{tabuchi2014hybridizing,zhang2014strongly,hyde2017linking,goryachev2014high,tabuchi2015coherent,lambert2016cavity}. Some interesting propositions involve multiple yttrium iron garnet (YIG) spheres placed inside an electromagnetic cavity, such as the magnon gradient memory \cite{zhang2015magnon}, the long distance modification of spin currents \cite{bai2017cavity}, and the development of ultrahigh sensitivity magnetometers\cite{PhysRevB.99.214415}.

As these new ideas are being elaborated \cite{hou2019strong}, it is important to correctly predict the behavior of photon-magnon systems consisting of several ferromagnetic elements coupled to an electromagnetic resonator. In this context, the effect of dipolar interactions between the ferromagnetic objects on the strong photon-magnon coupling is crucial and remains relatively unexplored.

The strong coupling of photon-magnon systems is well understood and has been recently reviewed\cite{harder2018cavity}. Its extension to multiple independent magnons is relatively straightforward\cite{zhang2015cavity}. For an ensemble of $N$ independent and identical ferromagnetic objects, the ideal coupling is expected to be enhanced by a factor of $\sqrt{N}$ as compared to the coupling strength of a single object to the cavity. However, due to dipolar interactions between the magnetic elements, some detuning along with inhomogeneous broadening are expected for coupled magnon systems. In this paper, we investigate the coupling strength of a simplified system consisting of two YIG spheres coupled to an adjustable microwave cavity. We show experimentally that the coupling constant $g$ decreases as the spheres are brought closer. The results are explained using a model based on the Landau-Lifshitz equation and the Fourier expansion of the magnetization in order to include the coupling of the photons with the uniform ferromagnetic mode as well as with the long wavelength spin wave modes, which are excited in presence of a non-uniform magnetic field.

\section{\label{sec:exp}EXPERIMENTAL PROCEDURE}
\begin{figure}[!t]
	\includegraphics[width=0.75\linewidth]{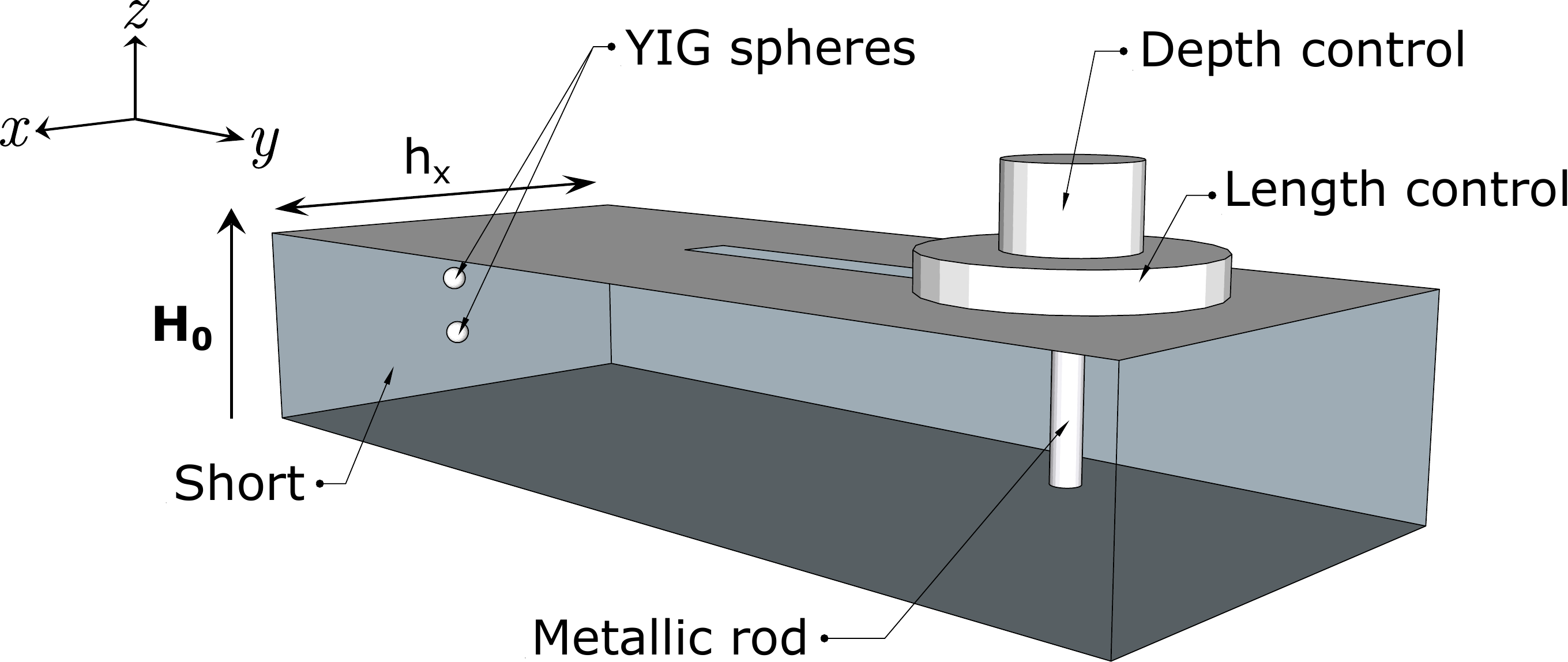}
\caption{\label{fig:tunecavity} Schematic representation of the tunable cavity used experimentally. The metallic rod allows the tuning of the resonance frequency and the losses of the cavity. The direction of the field $H_0$ and the RF magnetic field $h_x$ are also shown.}
\end{figure}

A tunable waveguide cavity consisting of a shorted X-band waveguide in which a metallic rod of $1.36$ mm of diameter is inserted in a slit located on one side of the waveguide \cite{morin2016strong} was used, as shown in Fig.~\ref{fig:tunecavity}.
Varying the position of the rod along the slit and its length inside the waveguide allowed the tuning of the volume, resonance frequency and electromagnetic losses of the cavity. For the experiments, the TE$_{109}$ mode with $\omega_c/2\pi=11.69$ GHz (volume $V_c=32.37$ cm$^3$) was chosen \cite{pozar2009microwave}. The cavity and the YIG spheres were excited by a vector network analyzer, which was also used as a detector to obtain the resonance spectra for different applied fields through the S$_{11}$ reflection coefficient. In order to observe the strong coupling regime, the spheres were placed on the shorted end of the waveguide resonator where the amplitude of the RF magnetic field is maximum. The two spheres, which will be called YIG$_1$ and YIG$_2$ hereafter, have a radius $R_1=0.62\pm0.01$ mm and $R_2=0.61\pm0.01$ mm, respectively. They were placed so that the center line (or axis) generated by the two spheres was parallel to the direction of the external DC magnetic field $H_0$. The center-to-center distance $d$ of the spheres was varied from $1.41$ mm to $3.58$ mm.

\begin{figure}[!t]
	\includegraphics[width=0.75\linewidth]{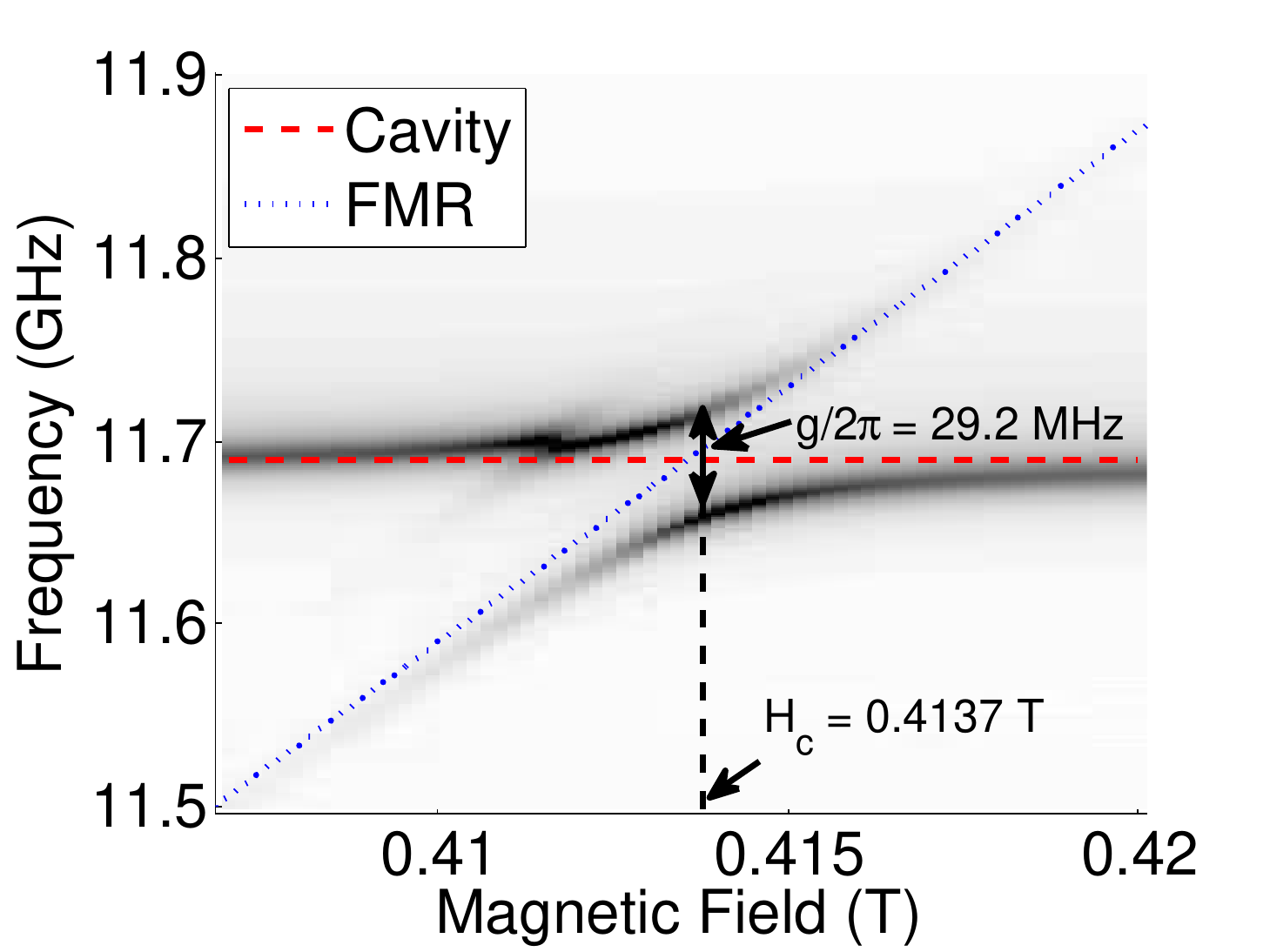}
\caption{\label{fig:YIG1} Strong coupling spectra obtained for the sphere YIG$_1$ with the setup described in Sect.~\ref{sec:exp}. The extracted coupling constant is $g_1/2\pi=29.2$ MHz. The hybridization of the modes occurs at a field $H_c=0.4137$ T. When there is no coupling, the resonance frequency of the cavity and the YIG sphere is represented by the red dashed line and the blue dotted line, respectively.}
\end{figure}

The strong coupling regime is observed when the coupling constant $g$ exceeds both the cavity losses $\kappa_c$ and the magnetic losses $\kappa_m$\cite{zhang2014strongly}. This is illustrated in Fig.~\ref{fig:YIG1} for YIG$_1$, where the hybridization of the cavity photonic mode and the ferromagnetic uniform mode of resonance is observed. In this work, the rod insertion was adjusted to have $\kappa_c$ comparable to $\kappa_m$ in order to facilitate the observation of the coupling. The coupling $g$ is obtained by subtracting the resonance frequency of both modes for the whole range of magnetic fields, whereas the minimum value is equal to $2g$. The value of the magnetic field corresponding to this minimum will be called $H_c$.

\section{\label{sec:res}RESULTS}
A coupling constant of $g_1/2\pi=29.2$ MHz and $g_2/2\pi=28.5$ MHz was independently extracted for YIG$_1$ and YIG$_2$, respectively. This agrees well with the theoretical value given by~\cite{tabuchi2014hybridizing}
\begin{equation}
g=\eta\sqrt{\frac{V_s}{V_c}}\left( \frac{\omega_M\omega_c}{2} \right)^{1/2},\label{constanteg}
\end{equation}
where $V_s$ is the volume of the sphere, $\omega_M = \mu_0|\gamma|M_s$ with $M_s$ the saturation magnetization, $\gamma$ the gyromagnetic ratio, and $\eta$ represents the spatial overlap between the cavity photonic mode and the magnonic mode. The factor $\eta$ is given by~\cite{lambert2015identification}
\begin{equation}
\eta = \left\rvert \frac{1}{h_{\text{max}}m_{\text{max}}V_s} \int_{\text{sphere}} (\mathbf{h}\cdot\mathbf{m})\text{d}V \right\rvert
\label{etafactor}
\end{equation}
where $\mathbf{h}$ is the dynamic magnetic field of the cavity with $h_{\text{max}}$ being its maximum magnitude and $\mathbf{m}$ is the dynamic magnetization of the sphere with $m_{\text{max}}$ being its maximum magnitude. The value of $\eta$ is usually equal to $1$ when $\mathbf{h}$ and $\mathbf{m}$ are both uniform, which is the case for a small sample placed at the maximum of the cavity field.

The input-output formalism \cite{zhang2014strongly} was used to extract the losses of each component. The losses of the cavity $\kappa_c/2\pi$ were $\approx 8.65$ MHz, similar to the losses $\kappa_{m1}/2\pi = 8.44$ MHz (YIG$_1$) and $\kappa_{m2}/2\pi = 12.63$ MHz (YIG$_2$) of the YIG spheres. The mean magnetic losses of both spheres, equal to $10.54$ MHz, will be referred to as $\bar{\kappa}_{m\infty}$ hereafter.
 
\begin{figure}[!t]
\centering
\subfloat{\includegraphics[width=1.7in]{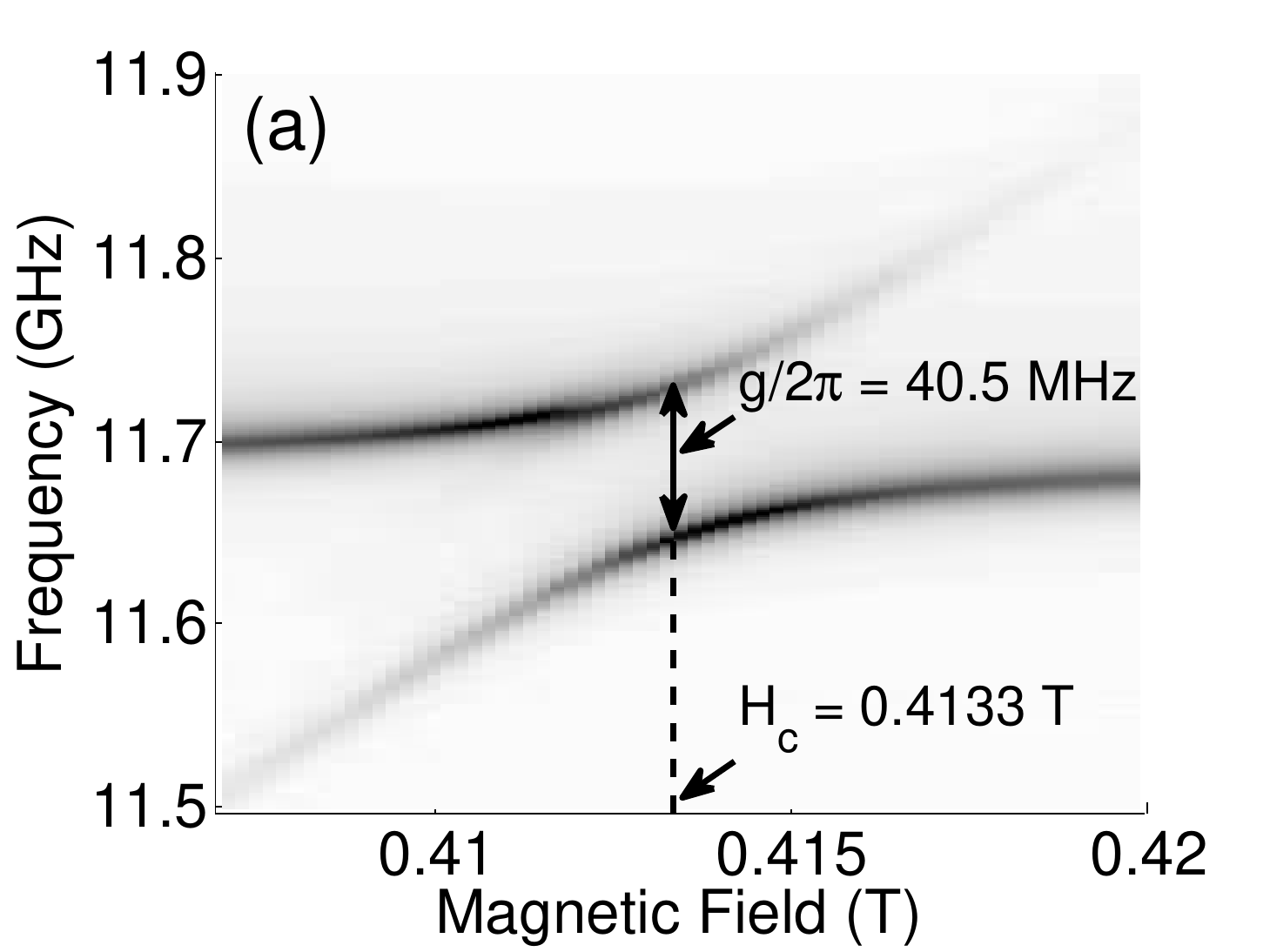}}
\hfil
\subfloat{\includegraphics[width=1.7in]{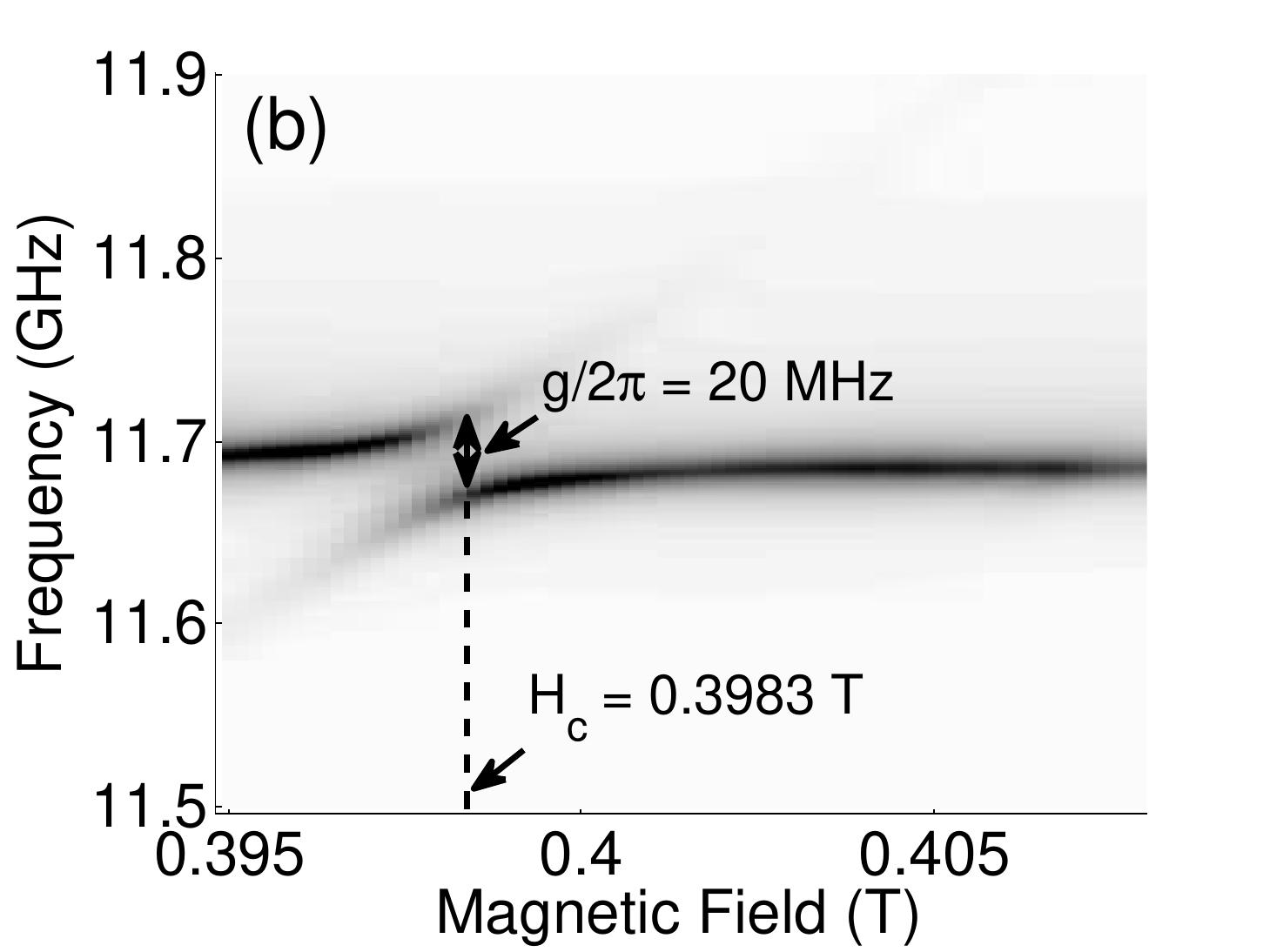}}
\caption{Strong coupling between the microwave cavity and two YIG spheres placed at a mutual distance of (a) $d = 3.58$ mm and (b) $d = 1.41$ mm. The dipolar interaction between the spheres shifts the value of $H_c$ and decreases the total coupling constant $g$.}
\label{fig:shift}
\end{figure}

With two spheres in the cavity, the hybridization of the modes is still exhibited, but accompanied with a shift in the value of $H_c$ and a change in the coupling strength as the spheres are brought closer. This is shown in Fig.~\ref{fig:shift} for two values of $d$. The field shift, due to the dipolar interaction, can be calculated by solving the coupled Landau-Lifshitz equations of motion of the two spheres treated as macrospins:
\begin{equation}
\frac{\partial}{\partial t} \left[ \begin{array}{c} \mathbf{M_1} \\ \mathbf{M_2} \end{array} \right] = -\mu_0|\gamma| \left[ \begin{array}{c} \mathbf{M_1} \\ \mathbf{M_2} \end{array} \right] \times \left( \mathbf{H}+\bar{\bar{N}}\left[ \begin{array}{c} \mathbf{M_2} \\ \mathbf{M_1} \end{array} \right] \right)\label{eq:LLG}
\end{equation}
where
\begin{equation}
\bar{\bar{N}} = \frac{1}{3}\left( \frac{R}{d} \right)^3 \begin{bmatrix} -1 & 0 & 0 \\ 0 & -1 & 0 \\ 0 & 0 & 2  \end{bmatrix},
\end{equation}
$\mathbf{H} = H_0\hat{z}+\mathbf{h}$, $R$ is the mean radius of the spheres, considered identical, and $d$ is the distance between the macrospins.
Using a small signal approximation, the coupled equations yield the resonance condition 
\begin{equation}
\omega_{\text{res}} = \omega_0 + \left( \frac{R}{d} \right)^3\omega_M
\label{rescond}
\end{equation}
where $\omega_0 = \mu_0|\gamma|H_0$. Because the hybridization of the modes occurs for $\omega_{\text{res}} = \omega_c$, Eq.~(\ref{rescond}) shows that smaller distances $d$ lead to smaller values of $H_c$. This shift of $H_c$ was used to corroborate and correct the distances between the spheres, which were initially measured manually with a digital micrometer. A good agreement has been found between the two methods. 

\begin{figure}[!t]
\includegraphics[width=0.75\linewidth]{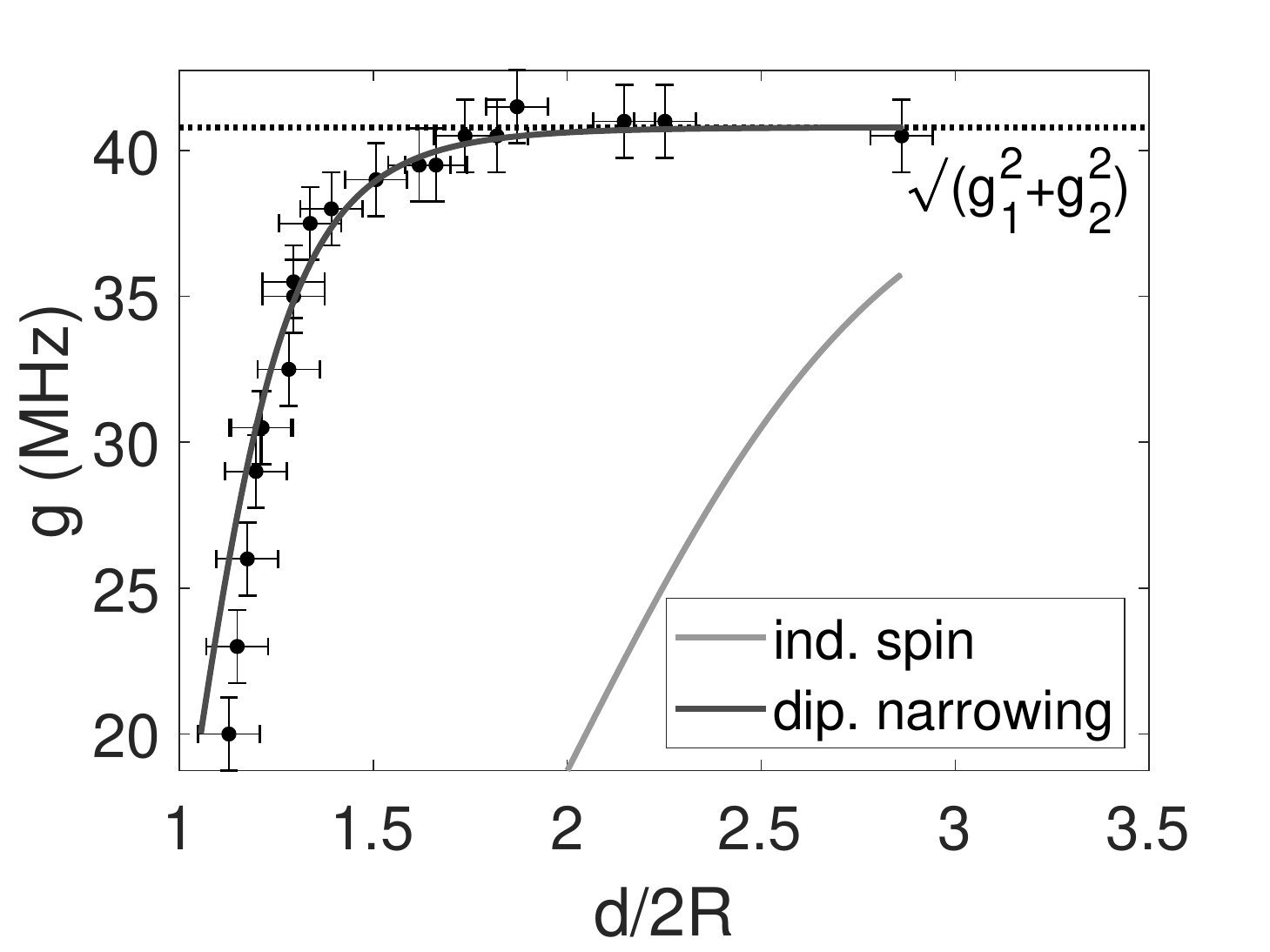}
\caption{\label{fig:dipolarint} Effect of dipolar interactions on the total coupling strength $g$ obtained experimentally (closed circles). Light gray curve: Expected decrease in the case of independent spins calculated with Eq.~(\ref{eq:ind}). Dark gray curve: Expected decrease in the case of dipolar narrowing calculated with Eq.~(\ref{eq:etaint}). Dotted line: Coupling constant when $d\rightarrow\infty$.}
\end{figure}

The reduction of the coupling constant $g$, exhibited in Figure~\ref{fig:shift} as the spheres are brought closer, is reported in greater details in Fig.~\ref{fig:dipolarint} (closed circles). For large distances between the spheres, one expects from the input-output formalism~\cite{schuster2010high} a total coupling strength of approximately $\sqrt{g_1^2+g_2^2}$ (dotted line), which is indeed observed. However, for smaller distances $d$, the coupling constant is observed to decrease sharply from $40.5$ MHz down to $20$ MHz.
\begin{figure}[!t]
\includegraphics[width=0.75\linewidth]{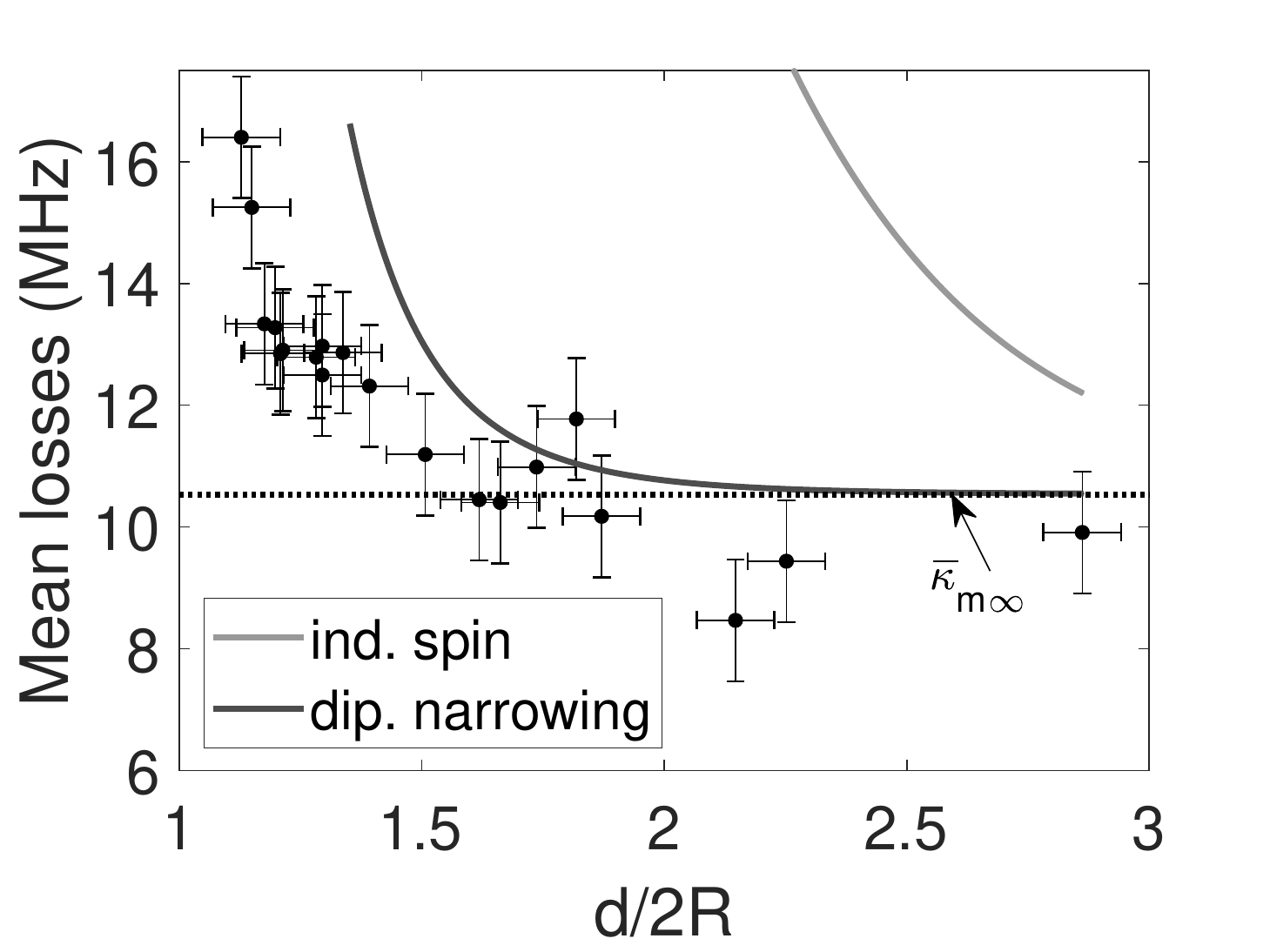}
\caption{\label{fig:losses} Magnetic losses $\bar{\kappa}_m$ obtained experimentally as the spheres get closer (closed circles). Light gray curve: Expected increase in linewidth when considering independent spins extracted from the susceptibility calculated with Eq.~(\ref{eq:ind}). Dark gray curve: $\bar{\kappa}_{m\infty}+\Delta\omega$ where $\Delta\omega$ is calculated using Eq.~(\ref{eq:delomega}). Dotted line: Magnetic losses when $d\rightarrow\infty$.}
\end{figure}

Considering the two YIG spheres as a whole, the usual expression of the S$_{11}$ reflection coefficient, calculated from the input-output formalism, was used to extract the magnetic losses of the two spheres as a function of the distance between the spheres $\bar{\kappa}_m(d)$ (closed circles in Fig.~\ref{fig:losses}). In contrast with $g$, the magnetic losses increase sharply as the spheres get closer. For a distance $d=1.41$ mm ($d/2R = 1.13$), the magnetic losses are just above $16$ MHz, which is near the coupling strength of $20$ MHz. For shorter distances, the magnetic losses would continue to increase while the coupling constant would decrease, causing the system to exit the strong coupling regime. 

\section{\label{sec:disc} DISCUSSION}
In order to explain the reduction of the coupling constant, let us consider the impact of the dipolar interaction on $\eta$. The dipolar field can be separated into two components. A dominant non-uniform static component is added to the applied static field and tend to spread the local field on the spheres. A weaker non-uniform dynamic field is further added to the cavity pumping field, which could result in the excitation of non-uniform resonance modes. Assuming that the RF magnetic field of the unperturbed cavity is uniform, we can rewrite $\eta$ in terms of the uniform mode susceptibility using $\mathbf{m} = \chi\mathbf{h}$. Since the real part of the susceptibility $\chi'\approx 0$ near resonance, we keep only the imaginary part and rewrite Eq.~(\ref{etafactor}) as
\begin{equation}
\eta = \frac{1}{\chi''_{\text{max}}V_s} \int_{\text{sphere}} \chi'' \ \text{d}V = \frac{\langle\chi''\rangle}{\chi''_{\text{max}}},\label{eq:etachi}
\end{equation}
where the brackets $\langle\cdot\rangle$ represents the mean value over the volume of the sphere.

Further insights are provided by examining two limiting cases. Case 1 corresponds to the macrospin approximation, in which all spins in a sphere are strongly coupled and locked parallel to each other's, which was assumed earlier in Eq.~(\ref{eq:LLG}). Our calculations indicate that the dynamic part of both spheres will be in phase, resulting in a constant factor $\eta=1$ for any distance $d$. In Fig.~\ref{fig:dipolarint}, the macrospin approximation corresponds to the dotted line and a value of $g = \sqrt{g_1^2+g_2^2}$. Likewise, the macrospin approximation does not lead to an increase in the linewidth observed in Fig.~\ref{fig:losses} but rather gives a constant linewidth of $\bar{\kappa}_{m\infty}$ (dotted line).

In contrast, Case 2 assumes fully independent spins, that is, no long-range dynamic dipolar interaction and each spin constituent of the spheres is resonating at its own frequency depending on the value of its local static magnetic field. This non-uniform magnetic field, assumed to be along the $\mathbf{\hat{z}}$-direction, is given by $H_z = H_0 + H_{\text{dip.}}$, where
\begin{equation}
H_{\text{dip.}} = \frac{R^3(r^2(3\cos^2\vartheta-1)+4dr\cos\vartheta+2d^2)}{3(r^2+2dr\cos\vartheta+d^2)^{5/2}}M_s.\label{eq:Hdip}
\end{equation}
Here, $H_{\text{dip.}}$ is the static dipolar magnetic field and the variables $r$ and $\vartheta$ determine the position in a spherical coordinates system centered on a sphere placed at a distance $d$ from the source dipole. One can then numerically compute the probability density function $f(H_{\text{dip.}})$ over the volume of the sphere as a function of $d$ to calculate the value of the mean susceptibility of the independent spins ensemble at resonance ($\omega=\omega_c$). Assuming no magnetic anisotropy, we have
\begin{equation}
\langle\chi''\rangle = \int_{H_z} \frac{\bar{\kappa}_{m\infty}\omega_M}{(\mu_0|\gamma|H_z-\omega_c)^2+\bar{\kappa}_{m\infty}^2}f(H_z) \text{d}H_z,\label{eq:ind}
\end{equation}
which can be substituted in (\ref{eq:etachi}) and then (\ref{constanteg}) to calculate the coupling. In this limiting case, a strong decrease in $\eta$ is predicted, even for spheres separated by a relatively large distance $d$, as shown by the light gray curve in Fig.~\ref{fig:dipolarint}. Furthermore, the inhomogeneously broadened linewidth in the independent spins approximation is given by the light gray line in Fig.~\ref{fig:losses}, which predicts a much broader linewidth than observed experimentally.

In our two spheres experiment, we thus fall somewhere between these two limits: macrospin and independent spins. A more rigorous approach should include long-range dynamic dipolar interactions which are known to produce a phenomenon called ``dipolar narrowing'' in the literature \cite{rezende1991dipolar}. We consider the original approach used by Clogston \cite{clogston1958inhomogeneous} in which the Landau-Lifshitz equation of motion is solved for a non-uniform magnetic field expanded in Fourier components as
\begin{align}
H_z = \sum_{\mathbf{k}} H_ke^{i\mathbf{k}\cdot\mathbf{r}}.
\end{align}
Assuming the field inhomogeneity is low with respect to the sample dimensions, we can neglect the terms related to the exchange interaction in the equation of motion, but consider the terms associated with dynamic dipolar fields. Further expanding the magnetization in Fourier series and by following a procedure similar to Ref.~\citenum{clogston1958inhomogeneous}, we can derive an analytical expression for the imaginary part of the susceptibility of the uniform mode of resonance, accounting for the coupling between the uniform mode and the long wavelength spin wave modes, a process called two-magnon scattering\cite{mcmichael2003localized}. With some simplifications, it can be written in the form
\begin{equation}
\langle \chi'' \rangle = \frac{(\bar{\kappa}_{m\infty}+\Delta\omega)\omega_M}{(\omega-\omega_c)^2+(\bar{\kappa}_{m\infty}+\Delta\omega)^2}\label{eq:chiClog}
\end{equation}
where
\begin{multline}
\Delta\omega = \frac{\pi}{2}\frac{\text{Var}(\omega_{\text{dip.}})}{\omega_M}\left[ 1+\frac{1}{2}\left( \frac{\omega_M}{3\omega_c-\omega_M} \right) \right]^2 \times \\ \left[ \frac{2}{3} - \left( \frac{\omega_c}{3\omega_c-\omega_M} \right) \right]^{-1/2}\label{eq:delomega}
\end{multline}
is an additional loss term directly related to the variance of the static dipolar magnetic field through the quantity $\omega_{\text{dip.}} = \mu_0|\gamma|H_{\text{dip.}}$, which can be calculated analytically (Appendix~\ref{sec:app1}). In the expression of $\Delta\omega$, the division by $\omega_M$ represents the dipolar narrowing effect. This additional loss term is added to $\bar{\kappa}_{m\infty}$, which yields a total loss term that can be compared with the measured mean losses of the magnetic system. As shown by the dark gray curve in Fig.~\ref{fig:losses}, the general trend of the data is reproduced relatively well.

Regarding the coupling constant $g$, the definition of $\eta$ in Eq.~(\ref{eq:etachi}) is extended to account for the fact that spins, whose resonance frequency $\mu_0|\gamma|H_z$ is detuned from the resonance frequency of the cavity $\omega_c$, can contribute to the coupling with the cavity. This can be achieved by introducing a weight function in the definition of $\eta$ so that the spins whose resonance frequency is contained inside the coupling range ($\pm g$ around $\omega_c$), have a stronger contribution (high energy exchange) to the total coupling than those whose resonance frequency falls outside the coupling range (low energy exchange). In contrast, in Eq.~(\ref{eq:etachi}), only the spins resonating at frequency $\omega_c$ contribute, whereas the remaining spins (detuned from the cavity) do not contribute to the coupling. To include this phenomenon, we use a weight function consisting in a Lorentz distribution $\mathcal{L}(\omega_c,g_{\text{max}})$ centered at $\omega = \omega_c$ and having a half-width at half maximum of $g_{\text{max}} = \sqrt{g_1^2+g_2^2}$. We thus have
\begin{equation}
\eta = \cfrac{\displaystyle\int_{0}^{\infty}\cfrac{(\bar{\kappa}_{m\infty}+\Delta\omega)\omega_M\mathcal{L}(\omega_c,g_{\text{max}})}{(\omega-\omega_c)^2+(\bar{\kappa}_{m\infty}+\Delta\omega)^2} \ \text{d}\omega}{\displaystyle\int_{0}^{\infty}\cfrac{\bar{\kappa}_{m\infty}\omega_M\mathcal{L}(\omega_c,g_{\text{max}})}{(\omega-\omega_c)^2+\bar{\kappa}_{m\infty}^2} \ \text{d}\omega},\label{eq:etaint}
\end{equation}
which equals unity if $\Delta\omega = 0$, in absence of dipolar broadening. Equation~(\ref{eq:etaint}) may be used with Eq.~(\ref{constanteg}) to generate the dark gray curve in Fig.~\ref{fig:dipolarint}. The excellent agreement with the experimental data supports that the observed decrease in the coupling rate between the system of magnetic spheres and the cavity, as the spheres are brought closer together, originates from the increasingly non-uniform dipolar static magnetic field on each sphere. It also shows that the long-range dynamic dipolar interaction within each sphere, which gives rise to the dipolar narrowing effect, somewhat limits the adverse effect of the non-uniform field distribution.

Similarly, the expression of $\eta$ given in (\ref{eq:etaint}) implies that a larger coupling $g_{\text{max}}$ tends to smooth out the adverse effect of a given dipolar broadening $\Delta\omega$ in reducing the total coupling strength.

\section{\label{sec:conc}CONCLUSION}
We have demonstrated that the dipolar interaction between two ferromagnetic objects can strongly affect their coupling with a microwave cavity. As the distance between the spheres is gradually reduced, dipolar interactions force the spins to resonate at increasingly different frequencies. This results in increased magnetic losses and decreased coupling strength $g$ of the system. A model based on inhomogeneous broadening with dipolar narrowing reproduces the main features observed on a system consisting of two YIG spheres in a tunable microwave cavity. While the reduction in the coupling strength can be linked with the variance of applied field caused by the dipolar interaction, this effect is attenuated by dipolar narrowing and by strong coupling of each individual sphere with the cavity.

Our results suggest that a number of $N$ individual ferromagnetic objects inserted in an electromagnetic cavity will eventually exhibit a reduced coupling as compared to the expected $g\propto \sqrt{N}$ behavior as the density is increased. Yet the dipolar broadening will be mitigated by a compensating dipolar narrowing effect. A trade-off must be found to determine the optimal density of ferromagnetic objects to be placed in the cavity to reach a maximum coupling strength while reducing the impact of dipolar interaction.

\appendix
\section{\label{sec:app1}Analytical expression for Var$(\omega_{\text{dip.}})$.}
Integrating by parts Eq.~(\ref{eq:Hdip}), we have
\begin{equation}
\langle\omega_{\text{dip.}}\rangle = \frac{a^3}{12} \omega_M
\end{equation}
where $a = 2R/d$ ($0\leq a\leq 1$). The integration by parts also leads to an analytical expression for $\langle\omega_{\text{dip.}}^2\rangle$. The definition of the variance, Var$(\omega_{\text{dip.}})=\langle\omega_{\text{dip.}}^2\rangle-\langle\omega_{\text{dip.}}\rangle^2$, then gives
\begin{widetext}
\begin{equation}
\frac{\text{Var}(\omega_{\text{dip.}})}{\omega_M^2} = \frac{a^3}{4} \left[ \frac{a}{3(4-a^2)^3}\left( 5+\frac{a^2}{2}\left( 1+\frac{a^2}{8} \right) \right) +\frac{\text{tanh}^{-1}(a/2)}{24}-\frac{a}{4}\left( \frac{3}{32}+\frac{a^2}{9} \right) - \frac{3(4-a^2)}{512}\text{ln}\left( \frac{2+a}{2-a} \right) \right].
\end{equation}
\end{widetext}

\bibliography{dipolar_interaction}

\end{document}